\documentclass[runningheads]{llncs}
\usepackage{graphicx}
\usepackage{epstopdf}
\begin{document}
\title{Does a robot path have clearance c?}
%
%
\author{Ovidiu Daescu \and
Hemant Malik}
\authorrunning{Ovidiu Daescu and Hemant Malik}
%
\institute{University of Texas at Dallas, Richardson TX 75080, USA \\
\email{\{daescu, malik\}@utdallas.edu}}
\maketitle              
\begin{abstract}
Most path planning problems among polygonal obstacles ask to find a path that avoids the obstacles and is optimal with respect to some measure or a combination of measures, for example an $u$-to-$v$ shortest path of clearance at least $c$, where $u$ and $v$ are points in the free space and $c$ is a positive constant. In practical applications, such as emergency interventions/evacuations and medical treatment planning, a number of $u$-to-$v$ paths are suggested by experts and the question is whether such paths satisfy specific requirements, such as a given clearance from the obstacles. We address the following path query problem: Given a set $S$ of $m$ disjoint simple polygons in the plane, with a total of $n$ vertices, preprocess them so that for a query consisting of a positive constant $c$ and a simple polygonal path $\pi$ with $k$ vertices, from a point $u$ to a point $v$ in free space, where $k$ is much smaller than $n$, one can quickly decide whether $\pi$ has clearance at least $c$ (that is, there is no polygonal obstacle within distance $c$ of $\pi$). To do so, we show how to solve the following related problem:   
Given a set \textit{S} of \textit{m} simple polygons 
in $\Re^{2}$, preprocess \textit{S} into a data structure so that the polygon in \textit{S} closest to a query line segment $s$ can be reported quickly. 
We present an $O(t \log n)$ time, $O(t)$ space preprocessing, $O((n / \sqrt{t}) \log ^{7/2} n)$ query time solution for this problem, for any $n ^{1 + \epsilon} \leq t \leq n^{2}$. For a path with $k$ segments, this results in $O((n k / \sqrt{t}) \log ^{7/2} n)$
query time, which is a significant improvement over algorithms that can be derived from existing computational geometry methods when $k$ is small.

\keywords{Path Query  \and Polygonal obstacles \and Clearance \and Proximity queries.}
\end{abstract}
\section{Introduction}

Path planning problems among polygonal obstacles in the plane usually ask to find a path that avoids the obstacles and is optimal with respect to some measure or a combination of measures, for example a shortest $u$-to-$v$ path~\cite{x,y,z} or an $u$-to-$v$ shortest path of clearance at 
least $c$~\cite{w1,w2}, where $u$ and $v$ are points in the free space and $c$ is a positive constant. In some practical applications however, such as emergency interventions/evacuations and medical treatment planning, a number of $u$-to-$v$ (polygonal or circular arc) paths are suggested by experts and the question is whether such paths satisfy specific requirements, such as a given clearance from the polygonal obstacles. This is illustrated in Figure~\ref{fig1}.
In this paper we address the following path query problem: 

\begin{figure}[]
\centering
\includegraphics[scale=0.4]{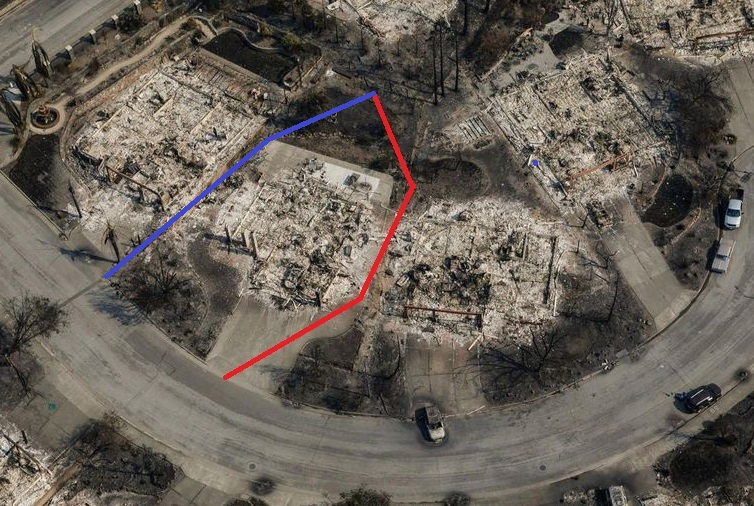}
\caption{California fire evacuation map, with a 4 feet clearance demand (drone acquired image). The shortest path (in blue) does not have good clearance. The proposed path (in red), should be checked (queried) for the desired clearance. }
\label{fig1}
\end{figure}

{\bf Path-Obstacles Proximity Queries}: Given a set $S$ of $m$ disjoint simple polygons in the plane, with a total of $n$ vertices, preprocess it to quickly answer queries of the following type: for a positive constant $c$ and a simple polygonal path 
$\pi$, from a point $u$ to a point $v$ in free space, decide whether $\pi$ has clearance at least 
$c$, that is, there is no polygonal obstacle within distance $c$ of $\pi$. 

Somehow surprisingly, it seems this problem has not been addressed in computational geometry. 
To solve it, we show how to solve the following related problem:
   
{\bf Object-Obstacles Proximity Queries}: Given a set \textit{S} of \textit{m} polygonal obstacles with a total of $n$ vertices, preprocess \textit{S} into a data structure so that the obstacle in \textit{S} closest to a query object $\rho$ can be reported quickly. 

In this paper we consider the set \textit{S} as a collection of disjoint simple polygons and the query object corresponds to a line segment (or line). 

Once the segment-polygon proximity problem is solved, one can check for each of the segments of the given path $\pi$ whether the segment has a clearance of $c$ or not, and also report the \emph{minimum clearance of the path}, defined as the minimum of the clearances of the line segments along the path.

\subsection{Related Work}
A simple, brute force solution to the path clearance problem would be to take each line segment along the path and find its distance (zero in case of intersections) to each of the line segments  
defining the boundary of the polygonal obstacles, which can be done in constant time per 
pair of segments. The clearance of the path would be reported as the minimum clearance over its line segments. For a path $\pi$ with $k$ line segments, among a set of
polygonal obstacles with a total of $n$ vertices, this leads to an $O(nk)$ time, $O(n+k)$ space solution that requires no preprocessing. This is linear in $n$ and thus inefficient 
for a query type problem. It is good however to contrast this with results that can be extracted from using complex data structures, such as the Visibility-Voronoi Complex 
(VVC)~\cite{w1}. The Visibility-Voronoi diagram for clearance $c$, $VV^{(c)}$, introduced in~\cite{w1}, encodes the visibility graph of the obstacles dilated with a disc of radius $c$ and can be used to compute paths of clearance $c$ and other desired properties between two points $u$ 
and $v$ by a search in this graph.
The Visibility-Voronoi complex is a generalization of $VV^{(c)}$, that allows to find $u$-to-$v$ paths for any given clearance value $c$ without having to first construct the $VV^{(c)}$, by performing a Dijkstra like search on the graph encoding the VVC. $VV^{(c)}$ and the VVC require
$O(n \log n + n_1)$ preprocessing time and can report an $u$-to-$v$ path of clearance at least $c$ in $O(n \log n + n_2)$ time, where $n_1$ is the number of visibility edges and $n_2$ is the number
of edges of the diagram encountered during the search; both $n_1$ and $n_2$ are $O(n^2)$ in the worst case. However, neither $VV^{(c)}$ or VVC can be used directly to check whether a given path 
has clearance at least $c$, since the edges of the path are in general not encoded by the underlying graphs.

For finding a closest point to a query line, Cole and Yap~\cite{cole} and Lee and Ching~\cite{lee}
reported a solution with preprocessing time and space in $O(n^2)$ and query time in $O(\log n)$.
Mitra and Chaudhuri~\cite{mitra} presented an algorithm with $O(n \log n)$ preprocessing time, 
$O(n)$ space, and $O(n^{0.695})$ query time.
Mukhopadhyay~\cite{muko}, used the simplicial partition technique
of Matousek~\cite{mat} to improve the query time to $O(n^{1/2+\epsilon})$) for arbitrary $\epsilon > 0$, with $O(n^{1+\epsilon})$) preprocessing time and $O(n \log n)$ space.

The problem of locating the nearest point to a query line segment
among a set $P$ of $n$ points in the plane was addressed in~\cite{besp}. 
If the query line segment is known to lie outside the convex hull of $P$, an $O(n)$ size data structure can be constructed in $O(n \log n)$ time, which can answer the nearest neighbor of a line segment in $O(\log n)$ time.
If $k$ non-intersecting line segments are given at a time, then the nearest neighbors of all these line segments can be reported in 
$O(k \log^3 n+ n \log^2 n+k \log k)$ time using divide and conquer and the data structure for queries outside the convex hull. 
Later on, in~\cite{besp1}, the time was reduced to $O(n \log^2n)$ when $n=k$. Moreover, 
given $n$ disjoint red segments and $k$ disjoint
blue segments in the plane, the algorithm in~\cite{besp1} can be used to find the closest pair of segments of a different color in $O((n+k) \log^2(n+k))$ time. Thus, with the red segments the edges of polygons in $S$ and the blue segments the segments along the query path, the path-polygon proximity problem we study can be solved within $O((n+k) \log^2(n+k))$ time, without any 
preprocessing. Our goal is to obtain a query time that is sublinear in $n$, and thus more efficient for small values of $k$. 

Goswami et al.~\cite{goswami} reported an algorithm for closest point to line segment queries with 
$O(\log^2n)$ query time and $O(n^2)$ preprocessing time and space, based on simplex range searching.
Segal and Zeitlin~\cite{Segal} provided an algorithm which takes $O(\log^{2} n \log \log n )$ 
query time, using $O(n^{2}/ \log n)$ space and $O(n^{2})$ preprocessing time. 
However, these algorithms do not answer the segment-polygon proximity query problem as described here, as it is not enough to consider only the vertices of the polygons in $S$. 

\subsection{Results}
For a set $S$ of $m$ disjoint simple polygonal obstacles in the plane, with a total of $n$ vertices, the goal is to preprocess $S$ so that given a positive constant $c$ and an $u$-to-$v$ polygonal path 
with $k$ edges, where $k$ is much smaller than $n$ (i.e., $k=o(n)$) one can quickly answer whether the path has clearance at least $c$.  
We have the following results.

\begin{itemize}
\item We present an $O(t \log n)$ time, $O(t)$ space preprocessing, 
$O((n / \sqrt{t}) \log ^{7/2} n)$ query time solution, for any $n ^{1 + \epsilon} \leq t \leq n^{2}$, to report the closest polygon in $S$ to a query line segment. 
\item For a path $\pi$ with $k$ segments, we obtain $O((n k / \sqrt{t}) \log ^{7/2} n)$
query time, with $O(t \log n)$ time, $O(t)$ space preprocessing, using segment-polygon proximity queries. When $t=n$ this gives $O(\sqrt{n} k \log ^{7/2} n)$ query time with linear space and 
$O(n \log n)$ time preprocessing, improving over previous methods whenever $k$ is small
(i.e. $k=o(\sqrt{n})$). When $t=\Theta(n^2)$ it gives a query time of $O(k \log ^{7/2} n)$, which is 
an $O(n)$ time improvement in query time over applying the solution derived from~\cite{besp,besp1}, when $k$ is small.
Moreover, unlike~\cite{besp,besp1}, our result is easily parallelizable, since queries with line 
segments along the query path are independent of each other. Thus, with $k$ processors available,
a query with a $k$ segment path would take time proportional to the time to answer a line segment query. Assuming $k$ is small, this can be easily implemented by multithreading (JAVA, C++) on modern laptop and desktop computers.
\end{itemize}


Our solutions differ from algorithms that could possibly be derived from existing visibility graph or Voronoi diagram based methods. They offer a preprocessing-query time
trade-off and result in significant improvements when $k$ is much smaller than $n$.

\section{Line segment proximity queries}  

\begin{figure}[]
\centering
\includegraphics[scale=0.35]{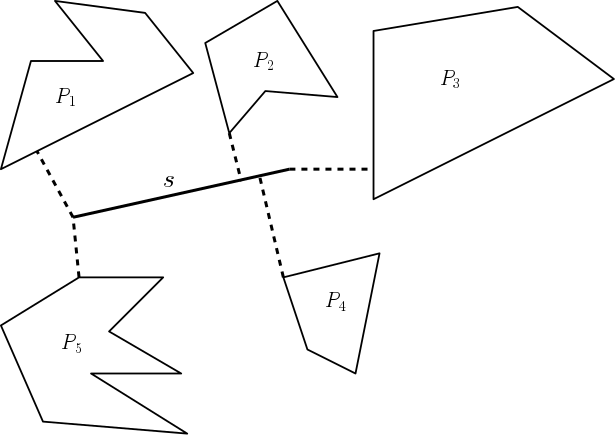}
\caption{A set \textit{S} of polygons, to be preprocessed for closest polygon to a query line segment (or line).}
\label{fig2}
\end{figure}

In a \textbf{nearest neighbor query} problem a set \textit{S} of \textit{n} geometric objects in $\Re^{d}$, $d$ a positive integer constant, is preprocessed into a data structure so that the object of \textit{S} closest to a query object (point, line, line segment, etc.) can be reported quickly. In this section we address nearest-neighbor queries in the plane, where the input  \textit{S} corresponds to a set of disjoint simple polygons and the query object corresponds to a line segment. This is illustrated in Figure ~\ref{fig2}. 

Obviously, if the query object intersects a polygon in $S$ then that polygon is a closest polygon and the closest distance from the query object to $S$ is zero. 
Following this observation, a query can be divided into two parts, executed in this order:

\begin{enumerate}
\item {\bf Emptiness Query}: Query if any polygon of $S$ is intersected by the query object. If there is such polygon, then report it as the answer, with a distance of zero. 

\item {\bf Proximity Query}: (No polygon in \textit{S} intersect the query object) Query for the closest polygon in $S$.  
\end{enumerate}

Thus, one can separately develop data structures for the two steps above, aiming for the best trade-offs on preprocessing-space-query on both structures.

Emptiness queries have been addressed in the context of ray shooting among polygonal obstacles in the plane.
Chazelle et al.~\cite{chazelle} gave an algorithm for ray shooting queries among $m$ disjoint 
simple polygons with a total of $n$ edges, with $O(n \sqrt{m} + m^{3/2} \log m + n \log n)$ preprocessing time, $O(\sqrt{m} \log n)$ query time, and $O(n)$ space.
Obviously, ray shooting queries can be used to answer emptiness queries for both lines and line segments, within the same time and space bounds, by replacing each such query with two, respectively one, ray shooting queries.
    
Agarwal and Sharir~\cite{Agarwal96} develop data structures for ray shooting queries by first
building data structures for line and line segment intersection queries. 
They first address line intersection queries and show that a set of $m$ simple polygons with a total of $n$ vertices can be preprocessed in time $O((m^2 + n \log m) \log n)$ into a data structure of size $O(m^2 + n)$ so that an intersection between a query line and the polygons can be detected on $O(\log n)$ time. Alternately, they give a data structure with $O(n \log n)$ preprocessing time, 
$O(n)$ space, and $O(\lceil m/\sqrt{n} \rceil ^{1+\epsilon} \log n)$ query time.
When $m \leq \sqrt{n}$ the query time becomes $O(\log n)$ while when $m \geq \sqrt{n}$ a query can 
be answered in time $O(\lceil m/\sqrt{t} \rceil ^{1+\epsilon})$ with space $t$ such that $n \leq t \leq m^2$. 
For line segment intersection queries for disjoint simple polygons they give a data structure of size 
$O((m^2+n) \log m)$, that can be constructed in $O((m^2+n) \log n \log m)$ time and 
can answer whether a query segment intersects any of the polygons in $O(\log m \log n)$ time.
Alternately, they gave a data structure with $O(n \log^2m)$ preprocessing time, $O(n \log m)$ space,
and $O(\lceil m/\sqrt{n} \rceil ^{1+\epsilon} \log^2n)$ query time.
For ray shooting among pairwise disjoint polygons, they give a data structure with 
$O(n \log n \log m)$ preprocessing time, $O(n)$ space, and $O(\lceil m/\sqrt{n} \rceil ^{1+\epsilon} \log^5n)$ query time.

We first warm up by providing a simple solution to finding the closest polygon to a query line
in the following subsection. We then extend this approach to find the closest polygon to a query line segment.  

\subsection{Closest polygon to a query line}


Given a set \textit{S} of \textit{m} disjoint simple polygons, with a total of $n$ vertices, to find the closest polygon to a query line $l$ we first perform an emptiness query with $l$, as described earlier. Using the result in~\cite{Agarwal96}, this can be done with 
$O(n \log n)$ preprocessing time, $O(n)$ space, and $O(\lceil m/\sqrt{n} \rceil ^{1+\epsilon} \log n)$ query time.

\begin{observation}
\label{obs1}
Given a simple polygon \textit{P} and a line \textit{l} such that \textit{l} does not intersect \textit{P}, the closest point of \textit{P} from \textit{l} is a vertex of \textit{P}. 
\end{observation}


Assume that none of the polygons in $S$ intersect the query line $l$.
Based on Observation~\ref{obs1} 
to find the closest polygon in $S$ to the query line $l$ reduces to finding the closest point to a query line problem, where points corresponds to the vertices of the polygons in \textit{S}. 
We further preprocess $S$ by computing the convex hull 
of each polygon in $S$ and taking the vertices of the convex hulls as the set of points. This requires only an additional $O(n)$ time and storage. Thus, we have a set of $n' \leq n$ points,
where $n'$ could be much smaller than $n$ in practice.  

We can then use the results in~\cite{muko,mitra} for closest point to line queries. 
Putting things together we obtain the following result.



\begin{lemma}
A set \textit{S} of $m$ polygons, with a total of $n$ vertices, can be preprocessed in 
$O(n^{1+\epsilon})$ time into a data structure of size $O(n \log n)$, that can report the closest polygon to a query line in 
$O(n^{(1/2) + \epsilon} + \lceil m/\sqrt{n} \rceil ^{1+\epsilon} \log n)$ time, for arbitrary $\epsilon > 0$.
Alternately, with $O(n \log n)$ time preprocessing one can construct a data structure of size $O(n)$ 
that can report the closest polygon to a query line in $O(n^{0.695})$ time.

\end{lemma}

\subsection{Closest polygon to a query line segment}

Given a set \textit{S} of \textit{m} polygons, with a total of $n$ vertices, to find the closest polygon to a query line segment $s$ we first perform an emptiness query with $s$.
To facilitate that, we preprocess $S$ into a data structure for planar point location queries, which requires $O(n \log n)$ time and $O(n)$ space~\cite{pointLocation}. Given a segment $s$, we locate the endpoints of $s$ in this data structure, in $O(\log n)$ query time; if any of the two endpoints is inside a polygon in $S$ then we are done. 

If both endpoints of $s$ are in free space we proceed with a segment intersection query.
As described earlier, this can be done with
$O((m^2+n) \log m)$ space, $O((m^2+n) \log n \log m)$ preprocessing time, and 
$O(\log m \log n)$ query time or, alternately, with $O(\lceil m/\sqrt{n} \rceil ^{1+\epsilon} \log^2n)$ query time, $O(n \log^2m)$ preprocessing time and $O(n \log m)$ space.


Assume that the line segment $s$ does not intersect any polygon in $S$. Obviously, the closest distance from $s$ to $S$ is attained by a point on $s$ and a point on a line segment on the
boundary of some polygon in $S$.    

\begin{observation}
\label{obs2}
Consider a line segment $e$ on the boundary of some polygon in $S$. The closest distance between $s$ and $e$ is either along:
\begin{enumerate}
\item A line perpendicular to $s$ and passing through an endpoint of $e$, or
\item  A line perpendicular to $e$ and passing through an endpoint of $s$, or
\item A line joining an endpoint of $s$ and an endpoint of $e$.
\end{enumerate}  
\end{observation}

\begin{figure}[] 
\centering
\includegraphics[scale=0.16]{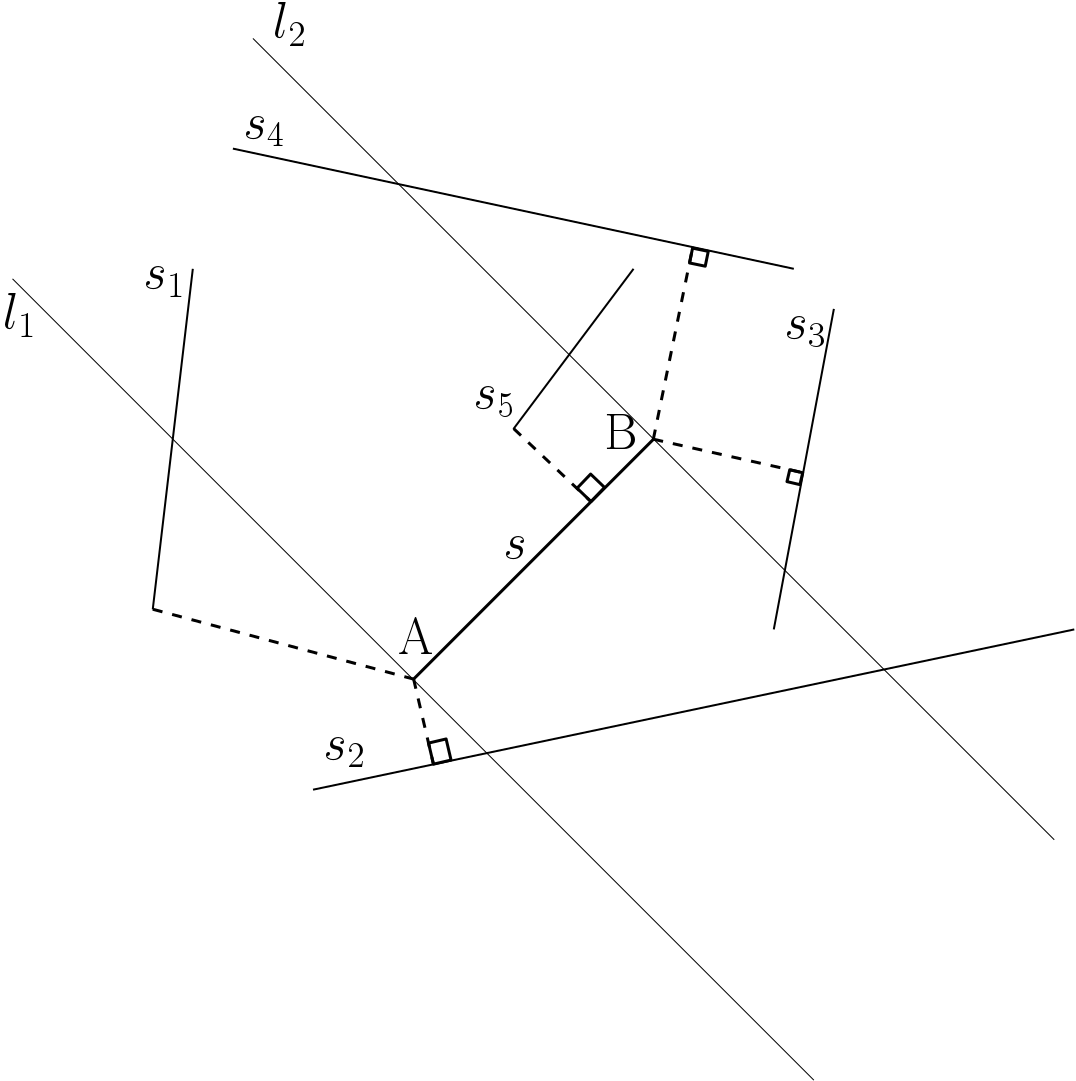}
\caption{Illustrating the closest distance between query line segment AB and some other line segments.}
\label{fig4}
\end{figure}

This is illustrated in Figure~\ref{fig4} where AB is the query segment $s$ and $s_{1}, s_{2}, s_{3}$, $s_{4}$ and $s_{5}$ are five line segments (see also~\cite{w1}). Lines $l_{1}$ and $l_{2}$ are perpendicular to segment $AB$ and passing through points \textit{A} and \textit{B}, respectively.

Thus, we can focus on the $n$ line segments on the boundaries of the polygons in $S$.  
Given a set $M$ of $n$ non-intersecting line segments, we want to build a data structure so that, for a query segment $s$ with endpoints $A$ and $B$, the closest segment of $M$ can be quickly determined. From Observation~\ref{obs2}, it is clear that the minimum distance involves an end point of at least one line segment. Therefore we can decompose this problem into the following two subproblems:

\begin{enumerate}
\item Find the line segment of set $M$ closest to point $A$ or point $B$.
\item Let $l_{1}$ and $l_{2}$ be the lines perpendicular to $s$ at its endpoints and consider the endpoints of the line segments in \textit{M} which lie between $l_{1}$ and $l_{2}$. Find the closest such endpoint to $s$.
\end{enumerate}


Notice that for our purpose we could relax the second subproblem, and ask instead for finding the closest endpoint of $M$ to the query segment $s$. 

\vspace{0.1in}
\noindent \textbf{Subproblem 1:} Given a set $M$ of $n$ line segments, preprocess $M$ so that we can efficiently find the closest line segment to a query point $q$. \\

To answer Subproblem 1 we construct the Voronoi diagram of the line segments in $M$ and preprocess it for
point location queries. Yap~\cite{Voronoi} provided an $O(n \log n)$ time algorithm to construct the Voronoi diagram of non intersecting line segments. After constructing the Voronoi diagram, 
preprocessing for point location takes $O(n \log n)$ time with $O(n)$ storage, and a point location query can be answered in $O(\log n)$ time~\cite{pointLocation}. 

\begin{lemma}
A set $M$ of $n$ non-intersecting line segments can be preprocessed in $O(n \log n)$ time  into a data structure of size $O(n)$ that can report the closest line segment to a query point in $O(\log n)$ time.
\end{lemma}

For a given line segment $s$, \textit{slab(s)} is defined as the region bounded by the lines 
$l_1$ and $l_2$ perpendicular to the endpoints of $s$ and containing $s$ (see~\cite{besp}).

\begin{figure}[]
\centering
\includegraphics[scale=0.26]{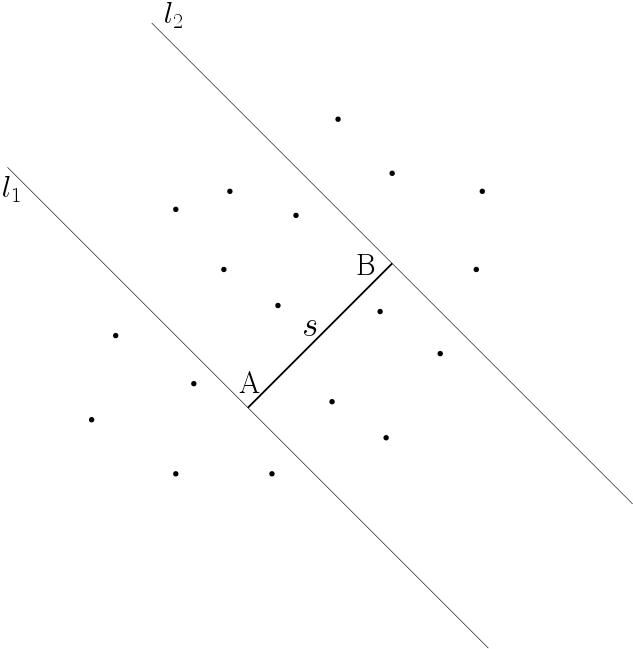}
\caption{slab(s) and the points in $P \in slab(s)$.}
\label{fig5}
\end{figure}

\vspace{0.1in}
\textbf{Subproblem 2:} Given a set $P$ of $n$ points, preprocess $P$ into a data structure so that one can efficiently answer the following query:  For a line segment $s$, find the closest point in $P \cap slab(s)$ to $s$. This is illustrated in Figure~\ref{fig5}.\\

To solve this problem, we use a multilevel data structure based on 
Matousek's~\cite{Matousek} decomposition scheme. Refer to Figure~\ref{fig5}. Specifically, the first level is for halfplane range queries, to separates the points that are on the side of $l_1$ that contains $s$, and the second level is for halfplane range queries on the resulting points to separate those that are on the side of $l_2$ and contain $s$. These two levels are used to isolate the points in $P \cap slab(s)$. The third level is for halfplane range queries bounded by the line supporting $s$, to isolate the subsets of $P \cap slab(s)$ that are on either side of $s$. The subsets on this level are further processed for closest point to line segment queries, when the query segment is outside the convex hull of the points, as in~\cite{besp}.

Using Theorem 6.1 from~\cite{Matousek}, we obtain the following trade-off.

\begin{lemma}
A set $P$ of $n$ points in the plane can be preprocessed in 
$O(t \log n)$ time and $O(t)$ space into a data structure that can answer the query in Subproblem 2 in $O((n / \sqrt{t}) \log^{7/2} n)$ time, for any $n ^{1 + \epsilon} \leq t \leq n^{2}$ and $\epsilon > 0$. 
\end{lemma}

Summing up, we have the following result.

\begin{lemma}
Given a set $S$ of $m$ disjoint simple polygons, it can be preprocessed in $O(t \log n)$ time into a data structure of size $O(t)$ that can report the closest polygon in $S$ to a query line segment in $O((n / \sqrt{t}) \log ^{7/2} n + \lceil m/\sqrt{n} \rceil ^{1+\epsilon} \log^2n)$ time, 
for any $n ^{1 + \epsilon} \leq t \leq n^{2}$ and $\epsilon > 0$. 
\end{lemma}

\section{Closest polygon to path queries}
We now turn our attention to finding the closest polygon to a query path.
Given a set $S$ of $m$ disjoint simple polygons in the plane, with a total of $n$ vertices, we want to preprocess $S$ into a data structure so that for a query consisting of a positive constant $c$ and a simple polygonal path $\pi$ with $k$ vertices, from a point $u$ to a point $v$ in free space, one can quickly decide whether there is no polygonal obstacle within distance $c$ of $\pi$.  

Our solution actually works even if $\pi$ has self intersections, however we do not see the practical aspect of such paths. 

To solve the path-polygons proximity query problem we proceed as follows.

\noindent {\bf Preprocessing.} We build the following data structures.
\begin{enumerate}
\item A point location data structure $D_1$ for the polygons is $S$. It can be built with $O(n)$ space and $O(n \log n)$ time, and can answer point location queries in $O(\log n)$ time.
\item A segment intersection data structure $D_2$ for the polygons of $S$.
As described earlier, this can be done with
$O((m^2+n) \log m)$ space, $O((m^2+n) \log n \log m)$ preprocessing time, and 
$O(\log m \log n)$ query time or, alternately, with $O(\lceil m/\sqrt{n} \rceil ^{1+\epsilon} \log^2n)$ query time, $O(n \log^2m)$ preprocessing time and $O(n \log m)$ space.
\item The Voronoi diagram of the polygons in $S$, enhanced with a point location data structure, $D_3$.
It can be built with $O(n)$ space and $O(n \log n)$ time, and can answer point location and closest polygon to point queries in $O(\log n)$ time.
\item  With $P$ the set of vertices of the polygons in $S$, a data structure $D_4$ to find the closest point of $P \cap slab(s)$ to $s$, for a query segment $s$, as described in the previous section.
It can be built with $O(t)$ space and $O(t \log n)$ time, and can answer a query in 
$O((n / \sqrt{t}) \log^{7/2} n)$ time, for any 
$n ^{1 + \epsilon} \leq t \leq n^{2}$ and $\epsilon > 0$.
\end{enumerate}

\noindent {\bf Query.} Given a simple polygonal path $\pi$ with $k$ vertices, to answer a query we proceed as follows. 
\begin{enumerate}
\item Query $D_1$ with the vertices of $\pi$. If any such vertex is inside some polygon of $S$
we stop and report it as the closest polygon to $\pi$, with a zero distance (or an 
{\em intersection} flag). Otherwise, all vertices of $\pi$ are in free space and we proceed with the next step. This step takes $O(\log n)$ time per query and thus $O(k \log n)$ time overall.
\item Query $D_2$ with the line segments on $\pi$. If it is found that a line segment intersects a polygon in $S$ then stop and report it as the closest polygon to $\pi$, with a zero distance (or an 
{\em intersection} flag).
\item Query $D3$ with the vertices of $\pi$ and keep tract of the closest distance found. That 
distance gives the closest polygon of $S$ to the vertices of $\pi$. 
This step takes $O(\log n)$ time per query and thus $O(k \log n)$ time overall.
\item Query $D4$ to find the closest obstacle vertex in $slab(s)$, for each segment $s$ of $\pi$.
This takes $O((n / \sqrt{t}) \log^{7/2} n)$ time per query and 
$O((k n / \sqrt{t}) \log^{7/2} n)$ time overall, for any 
$n ^{1 + \epsilon} \leq t \leq n^{2}$ and $\epsilon > 0$.  
\end{enumerate}

\begin{theorem}
A set $S$ of $m$ disjoint simple polygons in the plane, with a total of $n$ vertices, can
be preprocessed in $O(t \log n)$ time into a data structure of size $O(t)$ so that given a 
query consisting of a positive value $c$ and a simple polygonal path $\pi$ with $k$ vertices one
can answer if $\pi$ has clearance at least $c$ in 
$O(k((n / \sqrt{t}) \log^{7/2} n + \lceil m/\sqrt{n} \rceil ^{1+\epsilon} \log^2n))$ time, for any 
$n ^{1 + \epsilon} \leq t \leq n^{2}$ and $\epsilon > 0$.  
\end{theorem}

When $t=n^2$ and $m=o(\sqrt{n})$ the query time becomes $o(k \log^{7/2} n)$, which is a linear time faster than what can
be obtained from previous algorithms~\cite{besp,besp1} when $k$ is much smaller than $n$.
When $t=n^{1+\epsilon}$ the query time is 
$O(k(\sqrt{n} \log^{7/2} n + \lceil m/\sqrt{n} \rceil ^{1+\epsilon} \log^2n))$, which is asymptotically 
faster than what can be obtained from previous algorithms~\cite{besp,besp1} when $k=o(\sqrt{n})$. 

\section{Conclusion and Extensions}

In this paper, we studied the problem of finding the closest polygon of a set \textit{S} of 
disjoint simple polygons to a query line segment or simple polygonal path. We proposed 
solutions that are significantly better in query time, when $k$ is small relative to $n$, than what could be obtained from existing, non-query based approaches. 
Since queries with line segments along the query path are independent of each other, 
our result is easily parallelizable: with $k$ processors available,
a query with a $k$ segment path would take time proportional to the time to answer a line segment query. When $k$ is small, this can be easily implemented by multithreading (JAVA, C++) on modern laptop and desktop computers.

A possible extension of our work, that we leave as an open problem, is to query with paths that 
are not polygonal, but formed of, or including, circular arcs. 
This version has direct applications in minimally invasive surgery, for instruments formed of circular tubes~\cite{Linguraru17}. This problem seems significantly harder, even when the 
clearance $c$ (diameter of the tube) is known in advance.
We sketch a possible approach here and underline the missing data structures needed to
address this version.

It is easy to see that the minimum distance between a circular arc $\sigma$ and a line segment $s$ could be achieved by a point $p \in \sigma$ and a point $q \in s$ neither of which is an endpoint of 
$\sigma$ or $s$.  

For the general version, with clearance given at query time one would need data structures for the following two types of queries: (1) circular arc intersection queries for disjoint polygons and
(2) circular arc proximity queries for disjoint polygons. So far, neither of these data structures have been described in the computational geometry literature. There are however data structures for
ray shooting queries among circular arcs~\cite{agarwal1991intersection}, so if $k$ is comparable to $n$ one could instead answer
ray shooting queries against $\pi$ at query time. Such method however seems inefficient.

Consider now the case when the clearance $c$ is known at preprocessing time.
As before, we have a  set $S$ of $m$ disjoint simple polygons in the plane, with a total of $n$ vertices. In addition, we also know the clearance $c$, given as a positive real value. 
A query consists of a path $\pi$ with $k$ circular arcs and asks whether $\pi$ has clearance 
at least $c$. 
To solve it, one can proceed as follows.

\noindent {\bf Preprocessing.} Build the following data structures.
\begin{enumerate}
\item Find the Minkowski sum of the obstacles in $S$ with a disk of radius $c$ and compute the union
$\Gamma$ of the resulting objects, which can be done with $O(n)$ space and 
$O(n \log^2n)$ time~\cite{w1}. The boundary of $\Gamma$ consists of both
line segments and circular arcs. Further process $\Gamma$ for point location queries, for an additional $O(n)$ space and $O(n \log n)$ time. A point location query in the resulting data structure $D_1$ can be answered in $O(\log n)$ time. 
\item Preporcess $\Gamma$ for circular ray shooting queries: given a circular arc $\sigma$ determine
the first line segment or arc on the boundary of $\Gamma$ hit by $\sigma$. Notice that the radius
for the query circular arc is known in advance. 
\end{enumerate}

Thus, in this case, we are dealing only with a special case of circular ray shooting queries among disjoint line segments and circular arcs, where the radius of all circular arcs given at preprocessing time is the same. Still, we are not aware of any data structure that can efficiently handle such queries.

%
%
%
%

\end{document}